\documentclass[preprint]{aastex}
\newcommand{\etal}{et al.}
\newcommand{\kms}{$\rm km\,s^{-1}$}
\shorttitle{SDSS DR1}
\shortauthors{Adelman-McCarthy \etal}
\begin{document}
\title{The First Data Release of the Sloan Digital Sky Survey}
\author{
Kevork Abazajian\altaffilmark{\ref{Fermilab}},
Jennifer K. Adelman-McCarthy\altaffilmark{\ref{Fermilab}},
Marcel A. Ag\"ueros\altaffilmark{\ref{Washington}},
Sahar S. Allam\altaffilmark{\ref{Fermilab}},
Scott F. Anderson\altaffilmark{\ref{Washington}},
James Annis\altaffilmark{\ref{Fermilab}},
Neta A. Bahcall\altaffilmark{\ref{Princeton}},
Ivan K. Baldry\altaffilmark{\ref{JHU}},
Steven Bastian\altaffilmark{\ref{Fermilab}},
Andreas Berlind\altaffilmark{\ref{Chicago}},
Mariangela Bernardi\altaffilmark{\ref{CMU}},
Michael R. Blanton\altaffilmark{\ref{NYU}},
Norman Blythe\altaffilmark{\ref{APO}},
John J. Bochanski Jr.\altaffilmark{\ref{Washington}},
William N. Boroski\altaffilmark{\ref{Fermilab}},
Howard Brewington\altaffilmark{\ref{APO}},
John W. Briggs\altaffilmark{\ref{Chicago}},
J. Brinkmann\altaffilmark{\ref{APO}},
Robert J. Brunner\altaffilmark{\ref{Illinois}},
Tamas Budavari\altaffilmark{\ref{JHU}},
Larry N. Carey\altaffilmark{\ref{Washington}},
Michael A. Carr\altaffilmark{\ref{Princeton}},
Francisco J. Castander\altaffilmark{\ref{Barcelona}},
Kuenley Chiu\altaffilmark{\ref{JHU}},
Matthew J. Collinge\altaffilmark{\ref{Princeton}},
A. J. Connolly\altaffilmark{\ref{Pitt}},
Kevin R. Covey\altaffilmark{\ref{Washington}},
Istv\'an Csabai\altaffilmark{\ref{Eotvos},\ref{JHU}},
Julianne J. Dalcanton\altaffilmark{\ref{Washington}},
Scott Dodelson\altaffilmark{\ref{Fermilab},\ref{Chicago}},
Mamoru Doi\altaffilmark{\ref{IoAUT}},
Feng Dong\altaffilmark{\ref{Princeton}},
Daniel J. Eisenstein\altaffilmark{\ref{Arizona}},
Michael L. Evans\altaffilmark{\ref{Washington}},
Xiaohui Fan\altaffilmark{\ref{Arizona}},
Paul D. Feldman\altaffilmark{\ref{JHU}},
Douglas P. Finkbeiner\altaffilmark{\ref{Princeton}},
Scott D. Friedman\altaffilmark{\ref{STScI}},
Joshua A. Frieman\altaffilmark{\ref{Fermilab},\ref{Chicago}},
Masataka Fukugita\altaffilmark{\ref{ICRRUT}},
Roy R. Gal\altaffilmark{\ref{JHU}},
Bruce Gillespie\altaffilmark{\ref{APO}},
Karl Glazebrook\altaffilmark{\ref{JHU}},
Carlos F. Gonzalez\altaffilmark{\ref{Fermilab}},
Jim Gray\altaffilmark{\ref{Microsoft}},
Eva K. Grebel\altaffilmark{\ref{MPIA}},
Lauren Grodnicki\altaffilmark{\ref{Chicago}},
James E. Gunn\altaffilmark{\ref{Princeton}},
Vijay K. Gurbani\altaffilmark{\ref{Fermilab},\ref{Lucent2}},
Patrick B. Hall\altaffilmark{\ref{Princeton},\ref{PUC}},
Lei Hao\altaffilmark{\ref{Princeton}},
Daniel Harbeck\altaffilmark{\ref{MPIA}},
Frederick H. Harris\altaffilmark{\ref{NOFS}},
Hugh C. Harris\altaffilmark{\ref{NOFS}},
Michael Harvanek\altaffilmark{\ref{APO}},
Suzanne L. Hawley\altaffilmark{\ref{Washington}},
Timothy M. Heckman\altaffilmark{\ref{JHU}},
J. F. Helmboldt\altaffilmark{\ref{NMSU}},
John S. Hendry\altaffilmark{\ref{Fermilab}},
Gregory S. Hennessy\altaffilmark{\ref{USNO}},
Robert B. Hindsley\altaffilmark{\ref{NRL}},
David W. Hogg\altaffilmark{\ref{NYU}},
Donald J. Holmgren\altaffilmark{\ref{Fermilab}},
Jon A. Holtzman\altaffilmark{\ref{NMSU}},
Lee Homer\altaffilmark{\ref{Washington}},
Lam Hui\altaffilmark{\ref{Fermilab}},
Shin-ichi Ichikawa\altaffilmark{\ref{NAOJ}},
Takashi Ichikawa\altaffilmark{\ref{TohokuU}},
John P. Inkmann\altaffilmark{\ref{Fermilab}},
\v{Z}eljko Ivezi\'{c}\altaffilmark{\ref{Princeton}},
Sebastian Jester\altaffilmark{\ref{Fermilab}},
David E. Johnston\altaffilmark{\ref{Chicago}},
Beatrice Jordan\altaffilmark{\ref{APO}},
Wendell P. Jordan\altaffilmark{\ref{APO}},
Anders M. Jorgensen\altaffilmark{\ref{LANL}},
Mario Juri\'{c}\altaffilmark{\ref{Princeton}},
Guinevere Kauffmann\altaffilmark{\ref{MPA}},
Stephen M. Kent\altaffilmark{\ref{Fermilab}},
S. J. Kleinman\altaffilmark{\ref{APO}},
G. R. Knapp\altaffilmark{\ref{Princeton}},
Alexei Yu. Kniazev\altaffilmark{\ref{MPIA}},
Richard G. Kron\altaffilmark{\ref{Chicago},\ref{Fermilab}},
Jurek Krzesinski\altaffilmark{\ref{APO},\ref{MSO}},
Peter Z. Kunszt\altaffilmark{\ref{JHU},\ref{CERN}},
Nickolai Kuropatkin\altaffilmark{\ref{Fermilab}},
Donald Q. Lamb\altaffilmark{\ref{Chicago},\ref{EFI}},
Hubert Lampeitl\altaffilmark{\ref{Fermilab}},
Bryan E. Laubscher\altaffilmark{\ref{LANL}},
Brian C. Lee\altaffilmark{\ref{LBL}},
R. French Leger\altaffilmark{\ref{Fermilab}},
Nolan Li\altaffilmark{\ref{JHU}},
Adam Lidz\altaffilmark{\ref{Fermilab}},
Huan Lin\altaffilmark{\ref{Fermilab}},
Yeong-Shang Loh\altaffilmark{\ref{Princeton}},
Daniel C. Long\altaffilmark{\ref{APO}},
Jon Loveday\altaffilmark{\ref{Sussex}},
Robert H. Lupton\altaffilmark{\ref{Princeton}},
Tanu Malik\altaffilmark{\ref{JHU}},
Bruce Margon\altaffilmark{\ref{STScI}},
Peregrine M. McGehee\altaffilmark{\ref{NMSU},\ref{LANL}},
Timothy A. McKay\altaffilmark{\ref{Michigan}},
Avery Meiksin\altaffilmark{\ref{Edinburgh}},
Gajus A. Miknaitis\altaffilmark{\ref{Washington}},
Bhasker K. Moorthy\altaffilmark{\ref{NMSU}},
Jeffrey A. Munn\altaffilmark{\ref{NOFS}},
Tara Murphy\altaffilmark{\ref{Edinburgh}},
Reiko Nakajima\altaffilmark{\ref{Penn}},
Vijay K. Narayanan\altaffilmark{\ref{Princeton}},
Thomas Nash\altaffilmark{\ref{Fermilab}},
Eric H. Neilsen, Jr.\altaffilmark{\ref{Fermilab}},
Heidi Jo Newberg\altaffilmark{\ref{RPI}},
Peter R. Newman\altaffilmark{\ref{APO}},
Robert C. Nichol\altaffilmark{\ref{CMU}},
Tom Nicinski\altaffilmark{\ref{Fermilab},\ref{Lucent1}},
Maria Nieto-Santisteban\altaffilmark{\ref{JHU}},
Atsuko Nitta\altaffilmark{\ref{APO}},
Michael Odenkirchen\altaffilmark{\ref{MPIA}},
Sadanori Okamura\altaffilmark{\ref{DoAUT}},
Jeremiah P. Ostriker\altaffilmark{\ref{Princeton}},
Russell Owen\altaffilmark{\ref{Washington}},
Nikhil Padmanabhan\altaffilmark{\ref{Princetonphys}},
John Peoples\altaffilmark{\ref{Fermilab}},
Jeffrey R. Pier\altaffilmark{\ref{NOFS}},
Bartosz Pindor\altaffilmark{\ref{Princeton}},
Adrian C. Pope\altaffilmark{\ref{JHU}},
Thomas R. Quinn\altaffilmark{\ref{Washington}},
R. R. Rafikov\altaffilmark{\ref{IAS}},
Sean N. Raymond\altaffilmark{\ref{Washington}},
Gordon T. Richards\altaffilmark{\ref{Princeton}},
Michael W. Richmond\altaffilmark{\ref{RIT}},
Hans-Walter Rix\altaffilmark{\ref{MPIA}},
Constance M. Rockosi\altaffilmark{\ref{Washington}},
Joop Schaye\altaffilmark{\ref{IAS}},
David J. Schlegel\altaffilmark{\ref{Princeton}},
Donald P. Schneider\altaffilmark{\ref{PSU}},
Joshua Schroeder\altaffilmark{\ref{Princeton}},
Ryan Scranton\altaffilmark{\ref{Pitt}},
Maki Sekiguchi\altaffilmark{\ref{ICRRUT}},
Uros Seljak\altaffilmark{\ref{Princeton}},
Gary Sergey\altaffilmark{\ref{Fermilab}},
Branimir Sesar\altaffilmark{\ref{Zagreb}},
Erin Sheldon\altaffilmark{\ref{Chicago}},
Kazu Shimasaku\altaffilmark{\ref{DoAUT}},
Walter A. Siegmund\altaffilmark{\ref{Hawaii}},
Nicole M. Silvestri\altaffilmark{\ref{Washington}},
Allan J. Sinisgalli\altaffilmark{\ref{Princeton}},
Edwin Sirko\altaffilmark{\ref{Princeton}},
J. Allyn Smith\altaffilmark{\ref{Wyoming},\ref{LANL}},
Vernesa Smol\v{c}i\'{c}\altaffilmark{\ref{Zagreb}},
Stephanie A. Snedden\altaffilmark{\ref{APO}},
Albert Stebbins\altaffilmark{\ref{Fermilab}},
Charles Steinhardt\altaffilmark{\ref{Princeton}},
Gregory Stinson\altaffilmark{\ref{Washington}},
Chris Stoughton\altaffilmark{\ref{Fermilab}},
Iskra V. Strateva\altaffilmark{\ref{Princeton}},
Michael A. Strauss\altaffilmark{\ref{Princeton}},
Mark SubbaRao\altaffilmark{\ref{Chicago}},
Alexander S. Szalay\altaffilmark{\ref{JHU}},
Istv\'an Szapudi\altaffilmark{\ref{Hawaii}},
Paula Szkody\altaffilmark{\ref{Washington}},
Lidia Tasca\altaffilmark{\ref{MPA}},
Max Tegmark\altaffilmark{\ref{Penn}},
Aniruddha R. Thakar\altaffilmark{\ref{JHU}},
Christy Tremonti\altaffilmark{\ref{JHU},\ref{Arizona}},
Douglas L. Tucker\altaffilmark{\ref{Fermilab}},
Alan Uomoto\altaffilmark{\ref{JHU}},
Daniel E. Vanden Berk\altaffilmark{\ref{Pitt},\ref{Fermilab}},
Jan Vandenberg\altaffilmark{\ref{JHU}},
Michael S. Vogeley\altaffilmark{\ref{Drexel}},
Wolfgang Voges\altaffilmark{\ref{MPIEP}},
Nicole P. Vogt\altaffilmark{\ref{NMSU}},
Lucianne M. Walkowicz\altaffilmark{\ref{Washington}},
David H. Weinberg\altaffilmark{\ref{OSU}},
Andrew A. West\altaffilmark{\ref{Washington}},
Simon D.M. White\altaffilmark{\ref{MPA}},
Brian C. Wilhite\altaffilmark{\ref{Chicago}},
Beth Willman\altaffilmark{\ref{Washington}},
Yongzhong Xu\altaffilmark{\ref{Penn}},
Brian Yanny\altaffilmark{\ref{Fermilab}},
Jean Yarger\altaffilmark{\ref{APO}},
Naoki Yasuda\altaffilmark{\ref{NAOJ}},
Ching-Wa Yip\altaffilmark{\ref{Pitt}},
D. R. Yocum\altaffilmark{\ref{Fermilab}},
Donald G. York\altaffilmark{\ref{Chicago},\ref{EFI}},
Nadia L. Zakamska\altaffilmark{\ref{Princeton}},
Idit Zehavi\altaffilmark{\ref{Chicago}},
Wei Zheng\altaffilmark{\ref{JHU}},
Stefano Zibetti\altaffilmark{\ref{MPA}},
Daniel B. Zucker\altaffilmark{\ref{MPIA}}
}

\altaffiltext{1}{
Fermi National Accelerator Laboratory, P.O. Box 500, Batavia, IL 60510
\label{Fermilab}}

\altaffiltext{2}{
Department of Astronomy, University of Washington, Box 351580, Seattle, WA
98195
\label{Washington}}

\altaffiltext{3}{
Department of Astrophysical Sciences, Princeton University, Princeton, NJ
08544
\label{Princeton}}

\altaffiltext{4}{
Center for Astrophysical Sciences, Department of Physics \& Astronomy, Johns
Hopkins University, Baltimore, MD 21218
\label{JHU}}

\altaffiltext{5}{
Department of Astronomy and Astrophysics, The University of Chicago, 5640 S.
Ellis Ave., Chicago, IL 60637
\label{Chicago}}

\altaffiltext{6}{
Department of Physics, Carnegie Mellon University, Pittsburgh, PA 15213
\label{CMU}}

\altaffiltext{7}{
Center for Cosmology and Particle Physics,
Department of Physics,
New York University,
4 Washington Place,
New York, NY 10003
\label{NYU}}

\altaffiltext{8}{
Apache Point Observatory, P.O. Box 59, Sunspot, NM 88349
\label{APO}}

\altaffiltext{9}{
Department of Astronomy
University of Illinois
1002 W. Green Street, Urbana, IL 61801
\label{Illinois}}

\altaffiltext{10}{Institut d'Estudis Espacials de Catalunya/CSIC, Gran Capita 2-4,
08034 Barcelona, Spain
\label{Barcelona}}

\altaffiltext{11}{
Department of Physics and Astronomy, University of Pittsburgh, 3941 O'Hara
St., Pittsburgh, PA 15260
\label{Pitt}}

\altaffiltext{12}{
Department of Physics, E\"{o}tv\"{o}s University, Budapest, Pf.\ 32,
Hungary, H-1518
\label{Eotvos}}

\altaffiltext{13}{Institute of Astronomy and Research Center for the Early Universe, School
of Science, University of Tokyo,
 2-21-1 Osawa, Mitaka, Tokyo 181-0015, Japan
\label{IoAUT}}

\altaffiltext{14}{
Steward Observatory, 933 N. Cherry Ave, Tucson, AZ 85721
\label{Arizona}}

\altaffiltext{15}{
Space Telescope Science Institute, 3700 San Martin Drive, Baltimore, MD
21218
\label{STScI}}

\altaffiltext{16}{Institute for Cosmic Ray Research, University of Tokyo, 5-1-5 Kashiwa,
 Kashiwa City, Chiba 277-8582, Japan
\label{ICRRUT}}

\altaffiltext{17}{
Microsoft Research, 455 Market Street, Suite 1690, San Francisco, CA 94105
\label{Microsoft}}

\altaffiltext{18}{
Max-Planck Institute for Astronomy, K\"onigstuhl 17, D-69117 Heidelberg,
Germany
\label{MPIA}}

\altaffiltext{19}{
Lucent Technologies, 2000 Lucent Lane, Naperville, IL 60566
\label{Lucent2}}

\altaffiltext{20}{Departamento de Astronom\'{\i}a y Astrof\'{\i}sica, 
Facultad de F\'{\i}sica, Pontificia Universidad Cat\'{o}lica de Chile,
Casilla 306, Santiago 22, Chile
\label{PUC}}

\altaffiltext{21}{
U.S. Naval Observatory, Flagstaff Station, P.O. Box 1149, Flagstaff, AZ
86002-1149
\label{NOFS}}

\altaffiltext{22}{
New Mexico State University, Department of Astronomy, P.O. Box 30001, Dept
4500, Las Cruces, NM 88003
\label{NMSU}}

\altaffiltext{23}{
US Naval Observatory, 3540 Mass Ave NW, Washington, DC 20392
\label{USNO}}

\altaffiltext{24}{
Remote Sensing Division, Code 7215, Naval Research Laboratory, 4555 Overlook Avenue SW, 
Washington, DC 20375
\label{NRL}}

\altaffiltext{25}{National Astronomical Observatory, 2-21-1 Osawa, Mitaka, Tokyo 181-8588,
Japan
\label{NAOJ}}

\altaffiltext{26}{Astronomical Institute, Tohoku University, Aramaki, Aoba, Sendai 980-8578,
Japan
\label{TohokuU}}

\altaffiltext{27}{
Los Alamos National Laboratory, Los Alamos, NM 87545
\label{LANL}}

\altaffiltext{28}{
Max Planck Institute for Astrophysics, Karl Schwarzschildstrasse 1,
D-85748 Garching, Germany
\label{MPA}}

\altaffiltext{29}{
Mt. Suhora Observatory, Cracow Pedagogical University, ul. Podchorazych 2,
30-084 Cracow, Poland
\label{MSO}}

\altaffiltext{30}{
CERN, IT Division, 1211 Geneva 23, Switzerland
\label{CERN}}

\altaffiltext{31}{
Enrico Fermi Institute, The University of Chicago, 5640 S. Ellis Ave.,
Chicago, IL 60637
\label{EFI}}

\altaffiltext{32}{
Lawrence Berkeley National Laboratory, One Cyclotron Rd.,
Berkeley CA 94720-8160
\label{LBL}}

\altaffiltext{33}{
Astronomy Centre, University of Sussex, Falmer, Brighton BN1 9QJ, United
Kingdom
\label{Sussex}}

\altaffiltext{34}{
Department of Physics, University of Michigan, 500 East University Ave., Ann
Arbor, MI 48109
\label{Michigan}}

\altaffiltext{35}{
Institute for Astronomy
Royal Observatory
Blackford Hill
Edinburgh EH9 3HJ
Scotland
\label{Edinburgh}}

\altaffiltext{36}{
Department of Physics, University of Pennsylvania, Philadelphia, PA 19104
\label{Penn}}

\altaffiltext{37}{
Department of Physics, Applied Physics, and Astronomy, Rensselaer
Polytechnic Institute, Troy, NY 12180
\label{RPI}}

\altaffiltext{38}{
Lucent Technologies, 2701 Lucent Lane, Lisle, IL 60532
\label{Lucent1}}

\altaffiltext{39}{Department of Astronomy and Research Center for the Early Universe, School
of Science, University of Tokyo,
 7-3-1 Hongo, Bunkyo, Tokyo 113-0033, Japan
\label{DoAUT}}

\altaffiltext{40}{
Joseph Henry Laboratories, Princeton University, Princeton, NJ
08544
\label{Princetonphys}}

\altaffiltext{41}{
School of Natural Sciences, Institute for Advanced Study, Einstein Drive,
Princeton, NJ 08540
\label{IAS}}

\altaffiltext{42}{
Physics Department, Rochester Institute of Technology, 85 Lomb Memorial
Drive, Rochester, NY 14623-5603
\label{RIT}}

\altaffiltext{43}{
Department of Astronomy and Astrophysics, the Pennsylvania State
University, University Park, PA 16802
\label{PSU}}

\altaffiltext{44}{University of Zagreb, 
Department of Physics, Bijeni\v{c}ka cesta 32, 
10000 Zagreb, Croatia
\label{Zagreb}}

\altaffiltext{45}{
Institute for Astronomy, 2680 Woodlawn Road, Honolulu, HI 96822
\label{Hawaii}}

\altaffiltext{46}{
University of Wyoming, Dept. of Physics \& Astronomy, Laramie, WY 82071
\label{Wyoming}}

\altaffiltext{47}{
Department of Physics, Drexel University, Philadelphia, PA 19104
\label{Drexel}}

\altaffiltext{48}{
Max-Planck-Institut f\"ur extraterrestrische Physik, 
Giessenbachstrasse 1, D-85741 Garching, Germany
\label{MPIEP}}

\altaffiltext{49}{
Department of Astronomy, Ohio State University, Columbus, OH 43210
\label{OSU}}

\begin{abstract}
The Sloan Digital Sky Survey has validated and made publicly
available its First Data Release.  This consists of 2099 square
degrees of five-band ($u\,g\,r\,i\,z$) imaging data, 186,240 spectra of galaxies,
quasars, stars and calibrating blank sky patches selected over 1360 square degrees of this area, and
tables of measured parameters from these data.  The imaging data go to
a depth of $r \approx 22.6$ and are photometrically and
astrometrically calibrated to 2\% rms and 100 milli-arcsec rms per
coordinate, respectively.  The spectra cover the range 3800--9200\AA,
with a resolution of 1800--2100.  Further characteristics of the data
are described, as are the data products themselves. 
\end{abstract}

\section{Introduction}

The Sloan Digital Sky Survey (SDSS) is a photometric and spectroscopic
survey, using 
a dedicated 2.5-m telescope at Apache Point Observatory in New Mexico, of many
thousands of square degrees of high Galactic latitude sky.  The scientific goals that
define the scope of the project (York \etal\ 2000) relate to large-scale
structure seen in the distribution of galaxies and quasars.  In addition to
addressing these issues, the survey data products are proving valuable for
many other astronomical problems, from asteroids to Galactic
structure, from rare types of white dwarf stars to the highest-redshift
quasars.  The validated data are being released at 
approximately annual intervals.  Each release includes sufficient information to
allow statistical analysis, e.g. measures of data quality and the
completeness of the source lists.  The first SDSS Data Release (DR1) amounts
to about 20\% of the total SDSS survey goal.  

\begin{figure}[t]\centering\includegraphics[width=12cm]{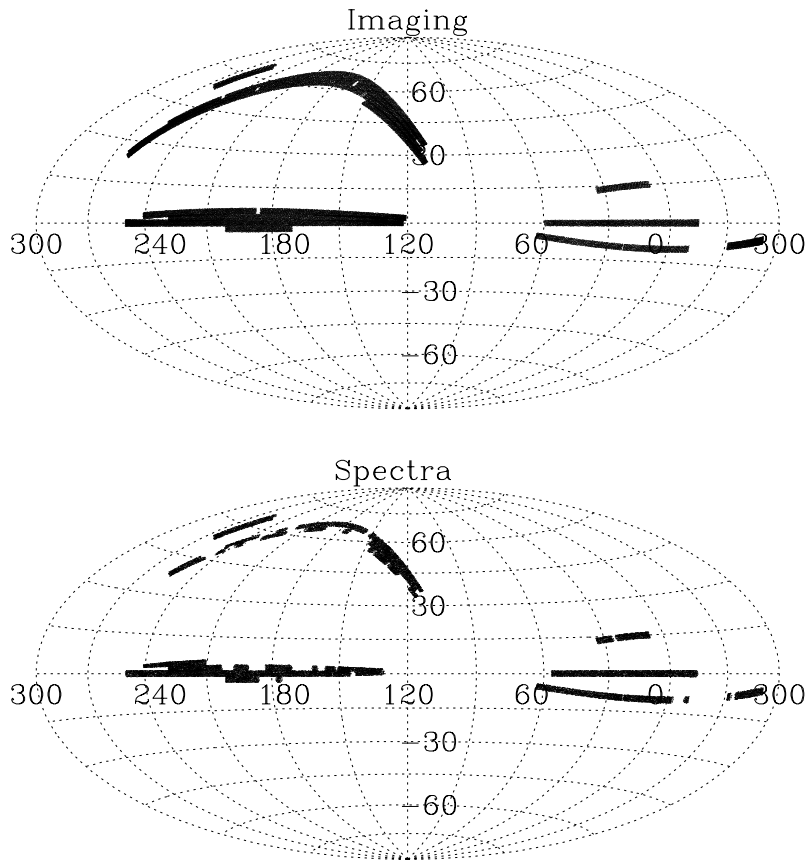}\caption{The
distribution on the sky of the imaging scans and spectroscopic plates
included in the DR1.  This is an Aitoff projection in equatorial
coordinates.  The total sky area covered by the imaging is
2099 square degrees, and by the spectroscopy is 1360 square degrees. 
\label{fig:dr1image}}\end{figure}

In Summer 2001, the SDSS released the results of observations obtained
during the commissioning phase of the SDSS; this Early Data Release
(EDR) is described by Stoughton \etal\ (2002), which contains
extensive information on the SDSS data and data processing software.  A
similarly comprehensive description of the DR1 data and derived
parameters may be found at {\tt http://www.sdss.org/dr1} (hereafter
the {\em web site}). 
The purpose of
the present paper is to formally mark the first SDSS data release and
to provide a quick guide to the contents of the {\em web site}. The sky coverage
of the imaging and spectroscopic components of the DR1 are shown in Figure~\ref{fig:dr1image}. 

\section{Published documentation}

A number of papers have been published that provide important technical
background relevant, but not limited, to DR1.  In this section we review
these publications.

A technical summary of the project is given by York \etal\ (2000).
This is an introduction to extensive on-line discussion of the
hardware (the Project Book), found at \\
{\tt http://astro.princeton.edu/PBOOK/welcome.htm}. The imaging camera is
described by Gunn \etal\ (1998).  

The Early Data Release is described by Stoughton \etal\ (2002), which
includes an extensive discussion of the data outputs and software.
More details of the photometric pipeline may be found in Lupton \etal\
(2001).  

\begin{figure}[t]\centering\includegraphics[width=12cm]{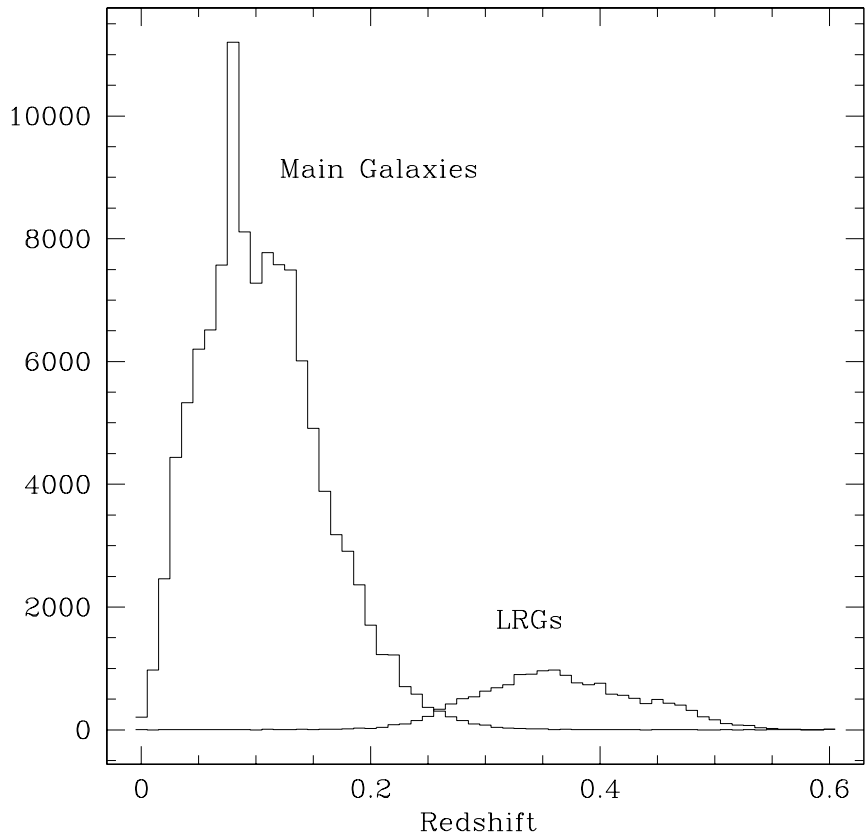}
\caption{Redshift histogram for objects spectroscopically classified
as galaxies in DR1.  The curve labeled ``Main Galaxies" is the
flux-limited sample, containing 113,199 galaxies.  The curve labeled
``LRGs" is a color-selected sample designed to contain intrinsically
luminous, red galaxies, containing 15,921 galaxies.
\label{fig:galaxydist}}\end{figure}

Strauss \etal\ (2002) give the target-selection procedures for the main
galaxy sample of the SDSS.  This paper provides the basis by which one can
construct a statistically complete sample of galaxies with spectra.
Eisenstein \etal\ (2001) describe the procedure for targeting 
a magnitude- and color-selected sample of Luminous Red Galaxies (LRGs)
at
redshifts up to $z=0.55$.   The redshift histograms of the objects
from these two samples in the DR1 are given in Figure~\ref{fig:galaxydist}.

\begin{figure}[t]\centering\includegraphics[width=12cm]{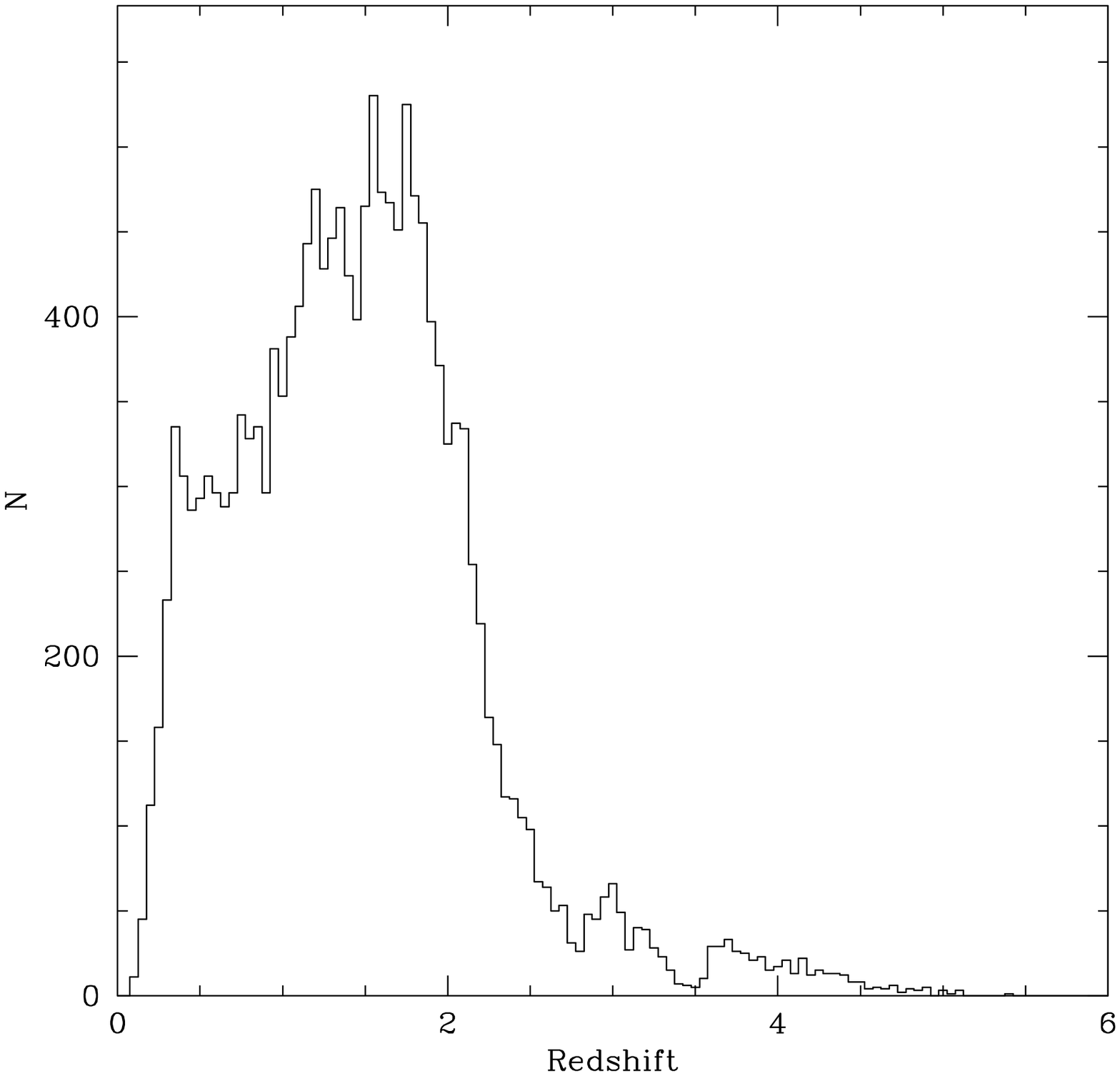}\caption{
Redshift histogram for objects spectroscopically classified as
quasars in DR1 (and with luminosities $M_B<-22$), including 16,847
objects.  The catalog of bona fide quasars in DR1 is
presented in Schneider \etal\ (2003). 
\label{fig:quasardist}}\end{figure}

Richards \etal\ (2002) present the algorithm that is
currently being used to target quasars from SDSS photometry, although
the DR1 sample (like the EDR sample) uses a more hetereogeneous set of
algorithms since the DR1 data predate the implementation of this
specific algorithm; see Schneider \etal\ (2003) for more details and a
formal catalog of DR1 quasars. The redshift histogram of spectroscopically
confirmed quasars in the DR1 is given in Figure~\ref{fig:quasardist}. 

These spectroscopic samples are assigned plates and fibers using an
algorithm described by Blanton \etal\ (2003). 

Pier \etal\ (2003) describe the methods and algorithms involved in the
astrometric calibration of the survey, and present a detailed analysis
of the accuracy achieved. 

The network of primary photometric standard stars is described by
Smith \etal\ (2002).  The photometric system itself is described by
Fukugita \etal\ (1996), and the system which monitors the site
photometricity is described by Hogg \etal\ (2001). 

The official (IAU designation) SDSS naming convention for an object is
SDSS~JHHMMSS.ss$\pm$DDMMSS.s, where the coordinates are truncated, not rounded.  This format should be
used at least once for every object listed in a paper using SDSS data.


\section{Contents of DR1}
The imaging portion of DR1 comprises 2099 square degrees of sky
imaged in five wavebands ($u, g, r, i$, and $z$), containing
photometric parameters of 53 million unique objects.  Within this
area, DR1 includes spectroscopic data (spectra and quantities derived
therefrom) for photometrically defined samples of quasars and
galaxies, as well as incomplete samples of stars.  The spectroscopic
data cover 1360 square degrees.  The details of the sky coverage can
be found at {\tt http://www.sdss.org/dr1/coverage/}. 

SDSS collects imaging data in strips which follow great circles.  Two
interleaving strips together make up a stripe 2.5 degrees wide; at the
equator of the system of great circles, stripes are separated by 2.5
degrees. 
A continuous scan of a piece of a
strip on a particular night is called a run; this is the natural
unit of imaging data.  Data from 62 runs are included in DR1.  The DR1
footprint is defined by all non-repeating survey-quality runs within the a priori
defined elliptical survey area (York \etal\ 2000) obtained prior to 1
July 2001; in fact, 34 square degrees of DR1 imaging data lie outside this
ellipse.  While the DR1 scans do not repeat a given area of sky,
they do overlap to some extent, and the data in the overlaps are
included in DR1 as well. 

Spectroscopy is undertaken with guided exposures of overlapping tiles
(``plates"), each 3 degrees in diameter.  For each plate, 640 spectroscopic
fibers are available.  DR1 consists of 291 plates, the centers of which lie
within the boundaries defined by the DR1 imaging footprint.

The surface density
of spectroscopic targets per square degree consists, on average, of roughly 90
galaxies in a flux-limited sample, an additional 12 galaxies of
a flux- and color-limited sample of Luminous Red Galaxies (LRGs), and
18 quasar candidates.  Each plate is assigned 18 calibration stars and
32 fibers on blank sky for sky subtraction.  Finally, extra fibers
available in a given region of sky are assigned to objects matching
ROSAT (X-ray; Voges \etal\ 1999) and FIRST (radio; Becker, White, \&
Helfand 1995) sources, as well as unusual stars of 
various types.  The spectroscopic targeting of all of these samples is
based on the photometric quantities produced by the SDSS pipelines.

DR1 includes the footprint of the sky already released in the EDR.
All of the EDR data have been reprocessed with the latest versions of
the SDSS software.  In some parts of the sky, better data (both
imaging and spectroscopy) have been
substituted for the older, commissioning data of the EDR.

The data products in DR1 include the following.

\begin{itemize} 
\item {\bf Images:} The ``corrected frames" (flat-fielded,
sky-subtracted, and calibrated sub-images corrected for bad columns,
bleed trails and cosmic rays, each $13.6' \times 9'$) in five
bands, available in both {\tt fits} and {\tt jpeg} format; a mask file
that records how each pixel was used in the imaging 
data-processing pipelines; $4 \times 4$-binned images (i.e., with
$1.6''$ pixels) of the corrected
frames after detected objects have been removed; and ``atlas images"
(cutouts from the corrected frames of each detected object).

\item {\bf Image parameters:} The positions, fluxes, shapes, and
errors thereof for all detected objects in the images, as well as
information about how these objects were spectroscopically targeted. 

\item {\bf Spectra:}  The flux- and wavelength-calibrated, sky-subtracted spectra, with
error and mask arrays, and the resolution of the spectra as a function
of wavelength. 

\item {\bf Spectroscopic parameters:}  The redshift and spectral
classification of each object with a spectrum, as well as the
properties of detected emission lines and various further spectral
indices. 

\item {\bf Other data products:}  Astrometric and photometric calibration files,
the point spread function of the images, {\tt gif} and postscript plots of
spectra, and ``finding charts" (cutouts of the survey image area
according to specified limits in right ascension and declination) in a
number of formats.  
\end{itemize}

These DR1 data products are available at the {\em web site} which
includes detailed description of the data, and documentation of the access tools.

\section{Changes in DR1 with respect to the Early Data Release}
\label{sec:changes}

The description of the SDSS data, file structures, and processing
pipelines presented in Stoughton \etal\ (2002) remains an essential
point of departure for understanding and using the data products in
the First Data Release.  However, there have been some significant
changes in the data processing since the EDR; we comment briefly here on some of the more
important ones.

\begin{itemize} 
\item The photometric equations have been reformulated to be in the
natural system of the 2.5-m telescope, making the relation between 
measured counts and magnitude a simple one.  The mean colors of stars on the
old and new systems have been forced to be the same.  The changes from
previously published photometry due to this are subtle, typically no
more than few hundredths of a magnitude.  To distinguish
between photometric systems, the new one ($u, g, r, i, z$) is unadorned,
whereas the EDR system was designated with asterisks ($u^*\,g^*\,r^*\,i^*\,z^*$).
The prime system discussed by Fukugita \etal\ (1996; $u'\,g'\,r'\,i'\,z'$)
now refers only to the native system of the US Naval Observatory
Flagstaff Station 1-meter telescope (cf., Smith \etal\ 2002), and should not be used in
referring to the data from the 2.5-m.   As before, all magnitude zero
points are approximately (i.e., within 10\%) on an AB system.  The magnitude scale is not
exactly logarithmic, but uses an asinh scaling (Lupton, Gunn \&
Szalay 1999; see the {\em web site} for further details).  Surface
brightnesses, however, are reported on a {\em linear} 
flux scale of ``maggies''; one maggie corresponds to the surface brightness
of a zeroth magnitude object in one square arcsecond.  A surface
brightness 20th mag in one square arcsecond is therefore $10^{-8}$
maggies. 

\item In the EDR, scattered light produced systematic errors in the derived flat
field, and therefore in the photometry, especially in the $u$ band.  
The imaging flat fields have now been corrected for this effect, reducing
a major source of systematic error.  

\item The EDR version of the photometric pipelines had difficulty following rapid
variations in the point-spread function.  The DR1 code is more robust
to this problem, and has greatly reduced the effects of variable
seeing on the photometric measurements. 

\item There is a small but measurable non-linearity in the response of
the photometric CCDs, measuring several percent at saturation. This
effect has been corrected in the DR1 processing.

\item The EDR image deblender often shredded galaxies with
substructure into several individual objects, especially for objects
brighter than $r \sim 15$ mag.  This behavior has been suppressed, and
the vast majority of bright galaxies are now treated properly by the
deblender.

\item In addition to the object shape measures of the EDR, the photometric
pipeline now calculates so-called adaptive moments (cf., Bernstein \&
Jarvis 2002) that are designed for
weak-lensing measurements of faint objects.

\item Cosmic rays are recognized as such by their sharp gradients relative to the
point-spread function.  An enhanced routine described in Fan \etal\ (2001)
is now implemented as part of the pipeline.  This routine sets a flag,
{\tt MAYBE\_CR}, which is valuable for assessing the reality of
objects detected in only a single band.  

\item In the EDR, the exponential (Freeman 1970) and de Vaucouleurs (1948) profile models for galaxy
images were fit only to the central 3 arcsec radius of each object. This
procedure tended to give misleading results for galaxies with large angular
extent. The DR1 version of the code does a much more reasonable fit to large
galaxies.  However, an error was found following the completion of DR1
processing, which causes the model magnitudes to be systematically
under-estimated by 0.2 magnitudes (i.e., the model magnitudes are too
{\em bright}) for galaxies brighter than 20th magnitude.  Similarly,
the measured radii are systematically too large.  This error will be corrected in
future releases.  Note that this error {\em only} affects model
magnitudes of galaxies; all other photometry is {\em unaffected} by
this error.  In addition, model {\em colors} are essentially
unchanged. 

\item The astrometric pipeline now uses centroids corrected for
asymmetries in the point-spread function, and includes a better treatment of chromatic
aberration. 

\item The spectroscopic pipeline has much improved flat-fielding, bias
subtraction, and handling of bad columns and pixels.  Sky subtraction has
been improved, especially in the red, by allowing for the gradient in the sky
brightness across a spectroscopic plate.  The spectrophotometric flux
calibration is improved as well, as is the correction for absorption lines
from the Earth's atmosphere.

\item There have been upgrades to the continuum and line-fitting
routines in the spectroscopic pipeline.  More extensive stellar
templates have increased the accuracy of the classification of unusual
types of stars.

\item The galaxy spectral-classification eigentemplates for DR1 are created from a
much larger sample of spectra (200,000) than were used for the EDR. 
\end{itemize}

\section{Data quality}

\subsection{Quality of Imaging Data}

\begin{figure}[t]\centering\includegraphics[width=12cm]{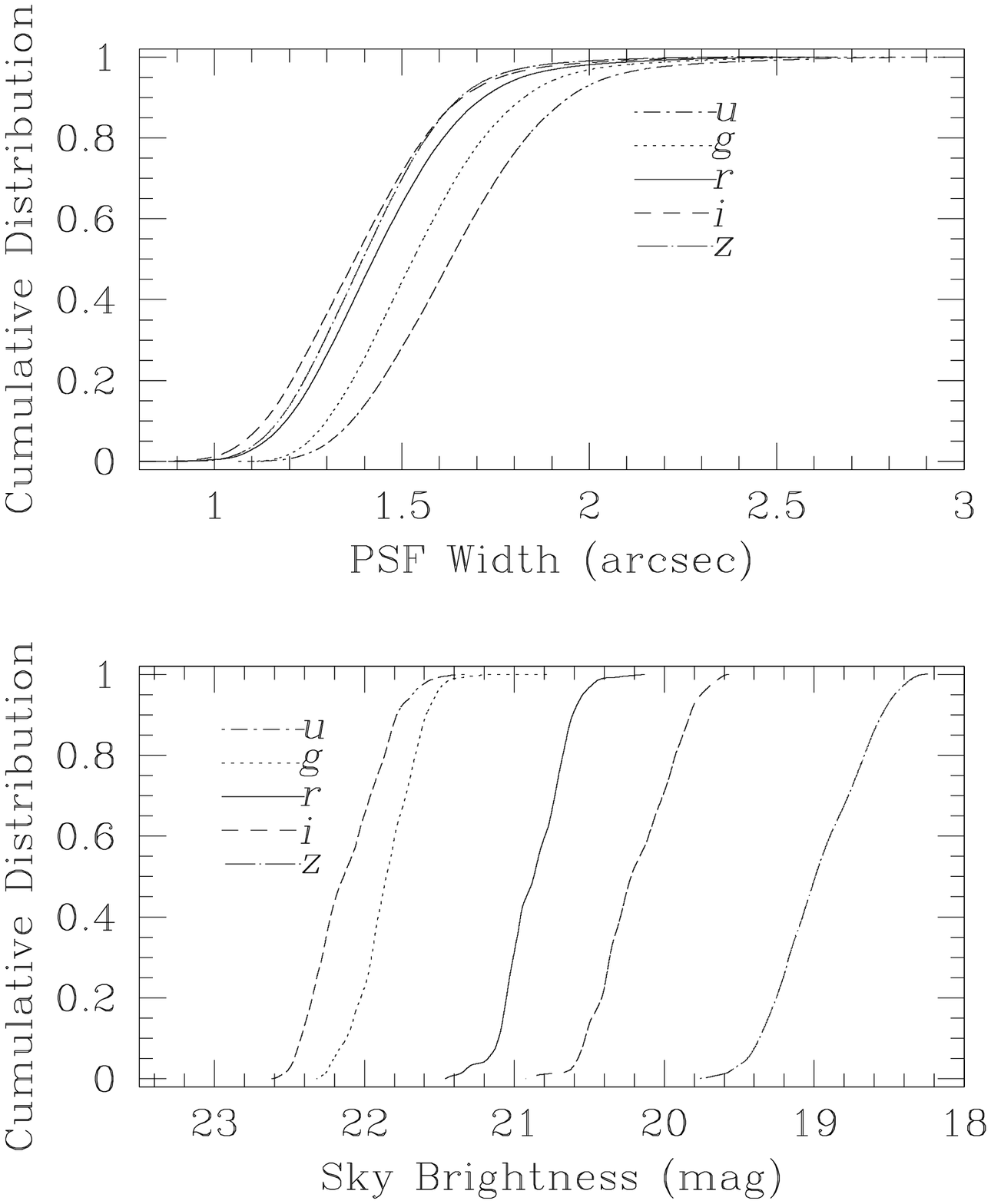}\caption{Top
panel: 
Cumulative
distribution
of values of the seeing 
in arcsec, for all frames in DR1, in each of the five
filters.  Note that over 90\% of the survey data meet the nominal
specification of seeing better than 1.7 arcsec in $r$.  Bottom panel:
Cumulative distribution of the values of the sky brightness in units of
magnitudes per square arcsec for all frames in DR1, as measured in the five
filters.
\label{fig:frames}}\end{figure}

The imaging survey is undertaken in photometric conditions (as
determined by an all-sky 10-micron camera) with no moon.  We also
impose a nominal limit on the effective width of the
point-spread function of 1.7 arcsec in the $r$ filter.  This width is
the full width at half maximum of the Gaussian with effective area
equal to that of the actual point spread function (PSF) at the center of each frame; it is
therefore somewhat larger than the actual full width at half maximum
of the PSF, due to the presence of extended low-amplitude wings on the
PSF.  In the
off-line processing, data are declared not to be of survey quality if the width of the
point-spread function exceeds this value for an interval longer than
about 10 minutes, or if the point-spread function is seen to be
rapidly varying, in which case an attempt is made to scan that
interval again.  The upper panel of Figure~\ref{fig:frames} shows the cumulative
distribution of the width of the point-spread function in DR1 as
determined on a frame-by-frame basis; only a very small fraction of
the data in DR1 exceeds the seeing threshold.  The five filters yield
different distributions both because of the dependence of seeing on
wavelength, and because the separate filters sample distinct regions
of the focal plane.

\begin{figure}[t]\centering\includegraphics[width=12cm]{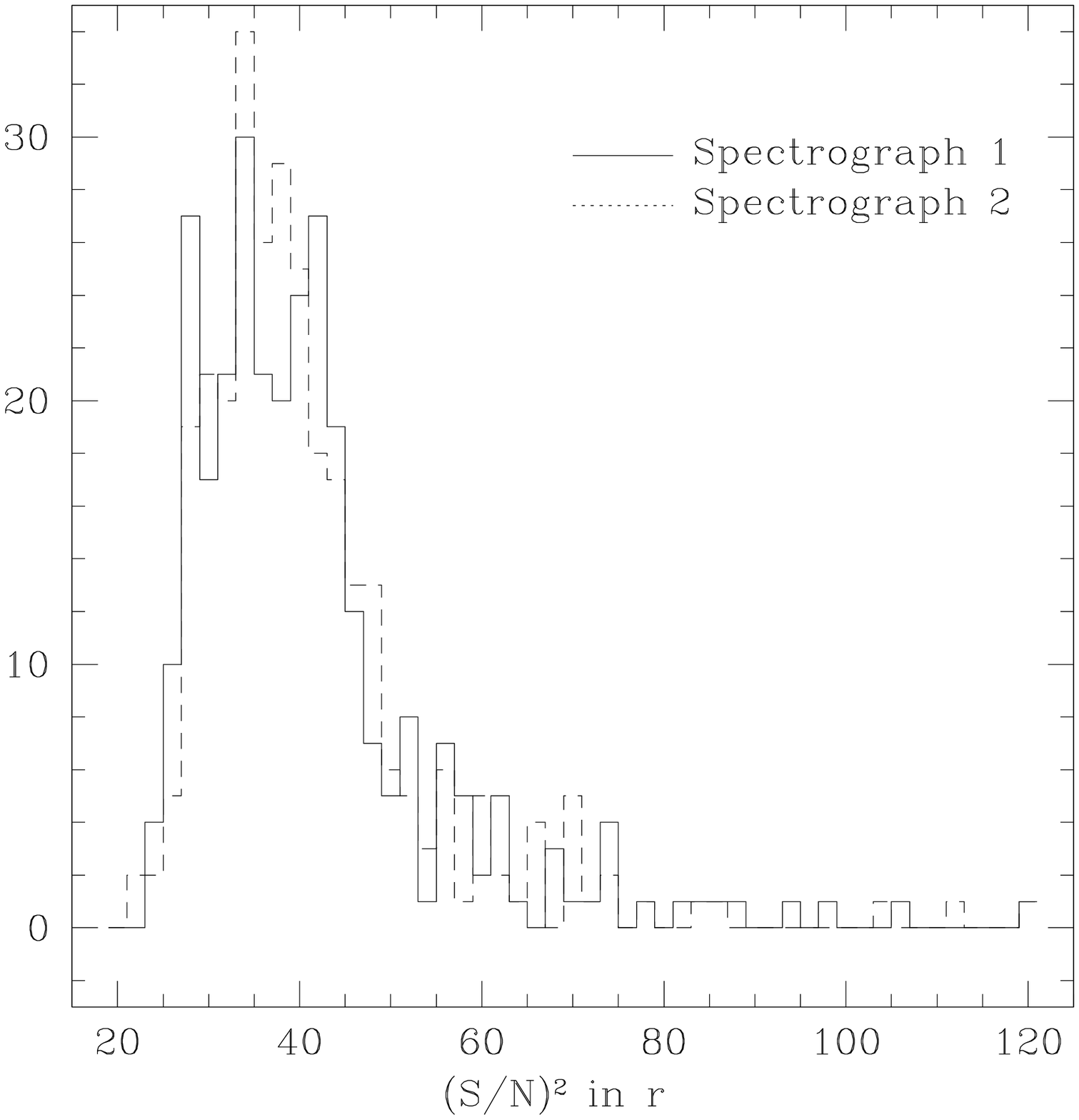}\caption{
Distribution of the values of the square of the spectroscopic
signal-to-noise ratio per pixel for an object with $r = 20.25$ through a
fiber,
 for the 291 plates in DR1.  The two spectrographs
(each with 320 fibers) are shown separately.  All the DR1 plates have
(S/N)$^2>20$; note the presence of a number of plates with signal to
noise ratio {\em much} higher than the minimum value. 
\label{fig:spectrasn}}\end{figure} 

The lower panel of Figure~\ref{fig:frames} shows the distribution of
sky brightness values in DR1, averaging over each frame.  This value
has been corrected for atmospheric extinction to zero airmass, and
therefore is biased to higher brightness, by 0.65 mags in $u$, but only 0.08 mag in
$z$.  The sky brightness, together with the instrumental throughput
and the atmospheric extinction and airmass,
allow one to compute the expected signal-to-noise ratio for the image
of an object of known brightness and profile; the sensitivity curves
for the five bands are available on the {\em web site}; cf., {\tt http://www.sdss.org/dr1/instruments/imager/}.

The photometric zero point is transferred from stars calibrated for this
purpose using an auxiliary Photometric Telescope, in $27 \times 27$
arcmin regions distributed approximately every 15 degrees 
along each stripe.  A number of tests allow us to quantify the 
uniformity of the photometric zero points and the accuracy of the
calibrations: 
\begin{itemize} 
\item Repeatability of photometry in regions of sky in which runs
overlap (cf., Ivezi\'c \etal\ 2003);
\item Constancy of the locus of stars in color-color space;
\item Lack of structure in the stellar or galaxy distribution on the
sky correlated with run geometry, seeing, foreground reddening, and
sky brightness;
\item Comparison of SDSS photometry with externally calibrated
standard star fields. 
\end{itemize}
From these results, the photometric zero point varies across the DR1
footprint by less than 0.02 mag rms in the $r$ 
band, 0.02 mag rms in the colors $(g-r)$ and $(r-i)$, and 0.03 mag rms in the
colors $(u-g)$ and $(i-z)$.

  We can similarly check the astrometric precision; repeat scans
confirm that our rms errors are rarely worse than 100 milli-arcsec
(mas) per coordinate; a more typical number is 60 mas.  See Pier
\etal\ (2003) for an extensive discussion of the astrometric accuracy.

The depth of the imaging data is a function of sky brightness and
seeing, but comparisons with deeper fields from the COMBO-17 survey
(Wolf \etal\ 2003) give 50\% completeness for
stellar sources at $(u,g,r,i,z) = (22.5,23.2,22.6,21.9,20.8)$ under
typical conditions.  Star-galaxy separation is better than 90\%
reliable to $r = 21.6$.

\subsection{Quality of Spectroscopic Data}
The spectroscopic survey is undertaken in observing conditions that
are not photometric, or with seeing worse than 1.7 arcsec FWHM, or
with some moonlight.  A spectroscopic observation is declared to be of
survey quality when the mean square of the signal-to-noise ratio per
pixel over the spectrum is greater than 15 for objects with fiber
magnitudes of $g=20.2$, $r=20.25$, and $i=19.9$; see the {\em web site} for
more details.  Figure~\ref{fig:spectrasn} presents the distribution of
the square of the signal-to-noise ratio per pixel for the 291 plates
in DR1, for objects with a fiber magnitude $r = 20.25$.  All the
plates clearly exceed the threshold of $(S/N)^2 = 15$; note the
presence of several plates with signal-to-noise ratio exceeding the
minimum requirements by factors approaching 3, as some plates were
observed for substantially longer times.  The sky subtraction is
sufficiently accurate that the noise is close to the photon shot noise.

 The FWHM of an unresolved emission line in the spectra is typically
2.5 pixels (1 pixel $\sim$ 65 \kms).  
From repeat observations of
galaxies near the survey limit, the redshift accuracy is known to be
better than 30 \kms; for bright stars, the redshift accuracy may be
better than 10 \kms.

The redshifts and classifications have been checked and updated by
comparing results from two independent codes.  Roughly 1\%
of the spectra (other than the 32 sky spectra per plate) are of low
enough signal-to-noise ratio as to be unclassifiable; of the
remaining, the error rate is below half a percent.

Data quality also depends on the precision and uniformity with which classes
of spectroscopic targets have been selected and observed.  The user
should be aware that the selection criteria for galaxies and quasars
have has ranged from 17.5 to 17.77 in extinction-corrected Petrosian
$r$ magnitude through the period covered by DR1.   In
particular, the magnitude limit for the main galaxy sample did move
about somewhat during the commissioning phase of the SDSS.  Similarly,
the quasar target selection algorithm described in Richards \etal\
(2002) is a modification of that used in DR1; the newer version is
more complete in high-redshift quasars.  See the
DR1 {\em web site} and the papers cited in Section 2 for more details.

\bigskip
    As the name implies, DR1 is the first of a series of releases of what
will eventually be the entire Sloan Digital Sky Survey.  The second data
release, DR2, is planned for early 2004.  DR2 will increase the total amount
of data by 50\% with respect to DR1, and it will include a reprocessing of
DR1 which will fix the model magnitude bug mentioned in \S~\ref{sec:changes}.

%
%
%




\acknowledgements
Funding for the creation and distribution of the SDSS Archive has been
provided by the Alfred P. Sloan Foundation, the Participating Institutions,
the National Aeronautics and Space Administration, the National Science
Foundation, the U.S. Department of Energy, the Japanese Monbukagakusho, and
the Max Planck Society. The SDSS Web site is {\tt http://www.sdss.org/}.

The SDSS is managed by the Astrophysical Research Consortium (ARC) for the
Participating Institutions. The Participating Institutions are The
University of Chicago, Fermilab, the Institute for Advanced Study, the Japan
Participation Group, The Johns Hopkins University, Los Alamos National
Laboratory, the Max-Planck-Institute for Astronomy (MPIA), the
Max-Planck-Institute for Astrophysics (MPA), New Mexico State University,
University of Pittsburgh, Princeton University, the United States Naval
Observatory, and the University of Washington.

\end{document}